\documentclass[english,a4paper]{article}
\usepackage[T1]{fontenc}
\usepackage[cp1250]{inputenc}
\usepackage{amssymb}
\usepackage{hyperref}
\usepackage{amsfonts}
\usepackage{graphicx}

\def\openone{\leavevmode\hbox{\small1\kern-3.3pt\normalsize1}}
\begin{document}

\title{Time optimization and state-dependent constraints in the quantum
  optimal control of molecular orientation}
%\date{\today}

\author{M. Ndong, C. Koch\footnote{Theoretische Physik, Universit\"at Kassel,
  Heinrich-Plett-Str. 40, 34132 Kassel,Germany} and D. Sugny\footnote{Laboratoire Interdisciplinaire
Carnot de Bourgogne (ICB), UMR 5209 CNRS-Universit\'e de
Bourgogne, 9 Av. A. Savary, BP 47 870, F-21078 DIJON Cedex,
FRANCE, dominique.sugny@u-bourgogne.fr}}

\maketitle

%\title{New formulation of quantum optimal control algorithms: \\

%  Time optimization and state-space constraints}

%%% it is not really a new formulation of the algorithms why not

%%%%%%%%%%%%%%%%%%%%%%%%%%%%%%%%%%%%%%%%%%%%%%%%%%%%%%%%%%%%%%%%%%%%%%%%

\begin{abstract}
We apply two recent generalizations of monotonically convergent
optimization
algorithms to the control of molecular orientation by laser fields.
We show how to minimize the control duration by a step-wise
optimization and maximize the field-free molecular orientation using
state-dependent constraints.
We discuss the physical relevance of the different results.
%%%%%%%%%%%%%%%%%%%%%%%%%%%%%%%%%%%%%%%%%%%%%%%%%%%%%%%%%%%%%%%%%%%%%%%%

%\bigskip

%\begin{keywords}

%Optimal control theory; molecular orientation; state-space constraints; time-optimization

%\end{keywords}\bigskip

\end{abstract}

%%%%%%%%%%%%%%%%%%%%%%%%%%%%%%%   INTRODUCTION   %%%%%%%%%%%%%%%%%%%%%%%%%%

\section{Introduction}

Optimal control tackles the question of bringing a dynamical system
from one state to another with minimum expenditure of time and
resources~\cite{pont,bryson}. When applied to quantum systems, control
is facilitated by matter wave interference, thus often termed
'coherent control'~\cite{RiceBook,ShapiroBook}. The target state is
reached by constructive interference while destructive interference
suppresses undesired outcomes. Optimal control has been applied
to quantum systems first in the context of physical
chemistry to steer chemical reactions~\cite{SomloiCP93,ZhuJCP98,revuerabitz},
followed by control of spin dynamics for applications in
NMR~\cite{SkinnerJMR03,grape}. Recently, optimal control is attracting much
attention in the context of quantum information processing, for example
as a tool to determine the minimum duration of high-fidelity quantum
gates~\cite{GoerzJPB11}, and for quantum simulation~\cite{DoriaPRL11}.

For quantum systems with complex dynamics and optimization targets
that are difficult to reach, it is impedient to utilize optimal
control algorithms that converge fast and monotonically.
The core of an optimization algorithm is made up of the control
equations which govern the system dynamics and update of the
control field. These equations are
derived by variation of the target and additional cost
functionals. Monotonicity can either be ensured by a smart
discretization of the coupled control equations~\cite{ZhuJCP98,gross,maday}
or built in using Krotov's
method~\cite{SomloiCP93,SklarzPRA02,PalaoPRA03}. The latter approach
comes with the advantage of independence from the specific form of
the optimization functional and matter-field
interaction~\cite{SklarzPRA02,ReichJCP12}. A non-linear matter-field
interaction is encountered in multi-photon couplings which are
important for example in the control of alignment and
orientation~\cite{nonlinear1,nonlinear2,NdongPRA13}. Additional
constraints in the optimization functional can be employed to keep
the system dynamics within a certain subspace~\cite{palao2008} or to
restrict the bandwidth of the optimized
field~\cite{WerschnikJOB05,GollubPRL08,SchroederNJP09,LapertPRA09,SkinnerJMR10}.
The control time~\cite{timeopt,zhang,timeopt1,timeopt2}
is also a crucial parameter and its optimization can help
to avoid, for example, parasitic phenomena with a longer time scale.

The purpose of the present paper is to test the efficiency of two
recent optimization procedures, namely state-dependent  constraints
and time-optimization, when applied to the orientation
dynamics of a linear molecule driven by an electromagnetic field. The
control of molecular alignment and orientation is by now a
well-recognized topic in quantum control with different applications
extending from the control of chemical reactions to nanoscale design
and quantum computing,
see Refs.~\cite{revuerotation2,revuerotation1} and references
therein. In the past few years, different methods have been proposed
to produce molecular orientation such as the use of non-resonant
light~\cite{friedrich,averbukh,salomon,ohtsukialign} or THz laser
pulses. The latter have the advantage to couple
resonantly to the molecular rotational
dynamics~\cite{fleischer,cong,lapert,henriksen}.
It is this option that will be investigated here.

The efficiency of the orientation is one important aspect in
view of applications. Another fundamental feature, which has received less
attention so far, is the time during which the molecular orientation
is above a given threshold under field-free conditions. A standard way to
maximize this duration is to restrict the dynamics to a given
subspace spanned by the lowest rotational
states~\cite{sugnyor1,sugnyor2}. Here we study the joint optimization of
orientation and its duration, employing the state-dependent constraint
algorithm of Ref.~\cite{palao2008}. Most of the theoretical and
experimental studies have so far been performed in the limit of an
isolated molecule where intermolecular collisions are
neglected~\cite{seideman,vieillard}. Dissipative effects such as those
due to collisions can be avoided if the optimization time is
sufficiently small. We investigate the question of optimization time
using the time-optimization
algorithm~\cite{timeopt,timeopt1,timeopt2}, which allows to find
the best compromise between the field fluence and the control duration.

The remainder of this paper is organized as follows. The
principles of monotonically converging optimal control
algorithms are outlined in Sec.~\ref{sec2}, with special attention
paid to state-dependent constraints and the time-optimization
formulation. Section~\ref{sec3} introduces the physical model.
The numerical results are presented and discussed in Sec.~\ref{sec4}.
We conclude in Sec.~\ref{sec5} with an outlook.

%%%%%%%%%%%%%%%%%%%%%%%%%%%%%%%%%%%%%%%%%%%%%%%%%%%%%%%%%%%%%%%%%%%%%

%%%%%%%%%%%%%%%%%%%%%%%%%%%%%%   SECTION 2   %%%%%%%%%%%%%%%%%%%%%%%%%%%%%%%%

\section{Review of optimal control algorithms}\label{sec2}
We present in this section three different optimal control algorithms,
considering pure quantum states and assuming the time evolution to be
coherent. The formalism is easily extended to mixed states or the
control of evolution operators by expanding in a
basis~\cite{PalaoPRA03}.  We take the  control target to maximize the
population of a target state,  but modification of the algorithms
to maximizing the expectation value of an observable is
straightforward. The dynamics of the  quantum system is governed by
the Hamiltonian
\begin{equation}
H(t)=H_0+E(t)H_1
\label{eq:Ham}
\end{equation}
with $E(t)$ the control field and $H_0$ the field-free
Hamiltonian. The operator $H_1$
describes the interaction between the system and the control field,
which we assume to be linear. We denote the initial and target states
by  $|\phi_0\rangle$ and $|\phi_f\rangle$, respectively, and represent
a general state by $|\psi(t)\rangle$.

\subsection{The standard formulation}
\label{subsec:stand}
We first review the standard formulation of optimal control
algorithms with the goal of bringing the system to a target
state~\cite{SomloiCP93,ZhuJCP98}.
The total time $t_f$ is fixed. The aim of the control problem is to
maximize the cost functional $J$,
\begin{equation}
  J[E]=|\langle \psi(t_f)|\phi_f\rangle |^2-\lambda
  \int_0^{t_f} [E(t)-E_{ref}(t)]^2/S(t)dt\,,
  \label{eq:cost}
\end{equation}
where $\lambda$ is a positive parameter which weights the relative
importance of the energy of the control field
with respect to the projection onto the target state. In
Eq. (\ref{eq:cost}), $E_{ref}(t)$ is a reference pulse  and $S(t)$
an envelope shape given by $S(t)= \sin^2(\pi t/t_f)$. The
function $S(t)$ ensures that the field is smoothly
switched on and off at the beginning and at the end of the control. We
consider a monotonic optimal control algorithm
which allows to increase the cost functional for any choice of the
free parameter of the system. More precisely,
we determine the field $E_{k+1}$ at step $k+1$ from the field $E_k$ at
step $k$, such that the variation $\Delta J=J(E_{k+1})-J(E_k)\geq 0$.
At step $k+1$, the reference field $E_{ref}(t)$ is taken to be
$E_k(t)$~\cite{PalaoPRA03}. Then the
correction of the control field at step $k+1$ is given by
\begin{equation}
  E_{k+1}=E_k+\frac{S(t)}{2\lambda}
  \mathfrak{Im} \left[\langle \chi_k |H_1|\psi_{k+1} \rangle \right]\,,
  \label{eq:newfield}
\end{equation}
where $|\chi_k (t)\rangle$ is obtained from backward propagation of
the target $|\phi_f \rangle$.
The dynamics of $|\chi_k(t)\rangle $ is governed by the Schr\"odinger
equation just as that for the state of the system
$|\psi_k(t)\rangle$.
The monotonic algorithm for the standard control problem of optimizing
state-to-state transfer is summarized as follows:
\begin{enumerate}
\item Guess an initial control field for $k=0$ and take
  $E_{ref}(t) = E_k(t)$ at step $k+1$.
\item Propagate forward in time the state of the system $|\psi_k\rangle$ with $E_k(t)$ from $|\phi_0\rangle$.
\item Starting from $|\phi_f\rangle\langle\phi_f |\psi(t_f)\rangle$, propagate $|\chi_k (t)\rangle$
  backward in time with $E_k(t)$.
\item Evaluate the correction of the control field, $E_{k+1}-E_k $,
  according to Eq.~(\ref{eq:newfield}),
  while propagating the state $|\psi_{k+1}\rangle$ forward  in time
  with $E_{k+1}$, starting from $|\phi_0\rangle$.
\item With the new control, $E_{k+1}$, go to step 3 by incrementing
  the index $k$ by 1.
\end{enumerate}

\subsection{State-dependent constraints}
\label{subsec:sdc}
Optimal control theory with a state-dependent constraint
\cite{palao2008} has been developed in order to
restrict the system evolution to a certain subspace.
This is useful in order to block multi-photon ionization that can be
caused by control fields with high intensity and to avoid subspaces
that are subject to decoherence. Formally, optimal control with a
state-dependent constraint is equivalent to optimizing a
time-dependent target~\cite{kaiser2004,serban2005}.
The corresponding cost functional is expressed as the sum of the
functional given in Eq.~(\ref{eq:cost}) with
a state-dependent intermediate-time cost \cite{palao2008},
\begin{equation}
  J^\mathrm{sdc}[E]=|\langle \psi(t_f)|\phi_f\rangle |^2-\lambda
  \int_0^{t_f} [E(t)-E_{ref}]^2/S(t)dt +
  \mu \int_0^{t_f} \langle \psi(t)| \hat{P} |\psi(t)\rangle dt\,,
  \label{eq:costsdc}
\end{equation}
where $\mu$ is  a positive weight parameter and $\hat{P}$ is here the
projector onto the allowed subspace. The control
equations  which allow for a monotonic increase of the functional,
Eq.~(\ref{eq:costsdc}), correspond to those obtained for the standard
algorithm modified by an inhomogeneity in the Schr\"odinger equation
for  $|\chi_k (t)\rangle$, the  backward propagated wave function,
\begin{equation}
  \frac{\partial}{\partial t} |\chi_k(t)\rangle=
  -i\left[H_0+H_1E_k(t)\right]|\chi_k(t)\rangle
  +\mu\hat{P}|\psi_k(t)\rangle\,.
  \label{eq:inho}
\end{equation}
Such an inhomogeneous Schr\"odinger equation can be solved e.g. by a
modified Chebychev propagator~\cite{ndong2009}.
At each iteration $k$, the new control field is given by
Eq.~(\ref{eq:newfield}) with the dynamics of the state $|\psi_k(t)\rangle$
governed by the homogeneous Schr\"odinger equation as before.
Other choices for the observable in the intermediate-time cost are
possible without perturbing monotonicity of the algorithm, provided the
operator is positive or negative semi-definite. This ensures that the
sign of the corresponding integral is well-defined~\cite{ReichJCP12}.

For a state-dependent constraint, the monotonic algorithm is
summarized as follows:
\begin{enumerate}
\item Guess an initial control field for $k=0$ and take
  $E_{ref}(t) = E_k(t)$ at step $k+1$.
\item Propagate the state of the system, $|\psi_{k}\rangle$, forward in
  time  with $E_{k}$, starting   from $|\phi_0\rangle$.
\item Evaluate the 'initial' condition, $|\chi(t_f) =
  |\phi_f\rangle\langle\phi_f |\psi(t_f)\rangle$, and
  propagate $|\chi_k (t)\rangle$ backward in time
  with $E_k(t)$ by solving the inhomogeneous Schr\"odinger equation,
  Eq.~(\ref{eq:inho})~\cite{ndong2009}.
\item Evaluate the correction to the control field, $E_{k+1}-E_k $,
  according to Eq.~(\ref{eq:newfield}),
  while propagating  the state  of the system, $|\psi_{k+1}\rangle$,
  forward in time with $E_{k+1}$, starting from $|\phi_0\rangle$.
\item With the new control, $E_{k+1}$, go to step 3 by incrementing
  the index $k$ by 1.
\end{enumerate}

\subsection{Time optimization}
\label{subsec:to}

Different formulations of monotonically convergent algorithms with time optimization have been proposed in
the literature \cite{timeopt,timeopt1,timeopt2}. Here we follow the approach introduced in Ref. \cite{timeopt}
where a cost penalizing both the field fluence and the control duration is used.
The cost functional to be maximized can be written as follows:
\begin{equation}
J^\mathrm{to}[E]=|\langle \psi(t_f)|\phi_f\rangle |^2-\lambda
\int_0^{t_f} [E(t)-E_{ref}(t)]^2/S(t)dt\,,
\end{equation}
with $t\in [0,t_f]$. A rescaling of time $s=t/t_f$ leads to normalized
quantities independent of $t_f$,
which will be denoted by a 'tilde' sign in the following. The new wave
function and control field are given by
\begin{equation}
|\tilde{\psi}(s)\rangle = |\psi (st_f)\rangle\,;\quad\tilde{E}(s)=E(st_f)\,,
\end{equation}
and the cost functional transforms into
\begin{equation}
\tilde{J}^\mathrm{to}[\tilde{E};t_f]=|\langle \tilde{\psi}(1)|\phi_f\rangle |^2-\lambda t_f
\int_0^{1} [\tilde{E}(s)-\tilde{E}_{ref}(s)]^2/\tilde{S}(s)ds\,.
\end{equation}
The goal of the algorithm is now to maximize $\tilde{J}^\mathrm{to}$
with respect to $\tilde{E}$ and the time $t_f$, which plays here the
role of a parameter.
%%%chr What do you mean the total time being fixed? is t_f not the total
%%%chr time? either it's a parameter or it's fixed
%Mamadou: J'ai enlever la phrase qui genait.

With this new time-parametrization,
the wave function $|\tilde{\psi}\rangle$ and the associated adjoint
state $|\tilde{\chi}\rangle$ satisfy
\begin{eqnarray*}
  i\frac{\partial}{\partial s}|\tilde{\psi} (s)\rangle &=&
  t_f [H_0+\tilde{E}(s)H_1 ] |\tilde{\psi}(s)\rangle \,,\\
  i\frac{\partial}{\partial s}|\tilde{\chi} (s)\rangle &=&
  t_f [H_0+\tilde{E}(s)H_1 ] |\tilde{\chi}(s)\rangle \,.
\end{eqnarray*}
Let us assume that at step $k$ of the iterative algorithm the system
is described by the quadruplet
$\left(|\tilde{\psi}_k\rangle, |\tilde{\chi}_{k}\rangle, \tilde{E}_k, t_f^{(k)}\right)$.
We determine the quadruplet at step $k+1$ from the one at step $k$ by
the following operations, which are decomposed into two substeps.
We first fix the time parameter $t_f^{(k)}$ and optimize the control
field by a standard algorithm,
as the one described in Sec.~\ref{subsec:stand}. We then get a new quadruplet
$\left(|\tilde{\psi}'_k\rangle, |\tilde{\chi}'_{k}\rangle,
  \tilde{E}_{k+1}, t_f^{(k)}\right)$.
In the second substep, we determine the new control time by keeping
the field $\tilde{E}_{k+1}$ fixed. A straightforward computation shows that the variation
$\Delta \tilde{J}^{\mathrm{to}}$ of the cost functional during this
step is given by
\begin{equation}\label{eqdj}
  \Delta \tilde{J}^{\mathrm{to}}= \left(t_f^{(k+1)}-t_f^{(k)}\right)
  \int_0^1 \mathfrak{Im}
  \left[\langle \tilde{\chi'}_k|H_{k+1}|\tilde{\psi}_{k+1}\rangle\right]ds
  -\lambda \left(t_f^{(k+1)}-t_f^{(k)}\right)\int_0^1
  \left(\tilde{E}_{k+1}-\tilde{E}_k\right)^2 ds\,,
\end{equation}
where $H_{k+1}=H_0+\tilde{E}_{k+1} H_1$. From Eq.~(\ref{eqdj}), it is
clear that the choice
\begin{equation}\label{eqtime}
  t_f^{(k+1)}=t_f^{(k)}+\varepsilon \left(\int_0^1 \mathfrak{Im}
    \left[\langle \tilde{\chi'}_k|H_{k+1}|\tilde{\psi}_{k+1}\rangle\right]ds
    -\lambda \int_0^1 \left(\tilde{E}_{k+1}-\tilde{E}_k\right)^2 ds\right)
\end{equation}
ensures an increase of $\tilde{J}^\mathrm{to}$. In Eq.~(\ref{eqtime}),
the new time $t_f^{(k+1)}$ is computed from
$|\tilde{\psi}_{k+1}\rangle$, whose propagation requires the value of
$t_f^{(k+1)}$.
A small parameter $\varepsilon$ such that $t_f^{(k+1)}\simeq
t_f^{(k)}$ allows to replace $t_f^{(k+1)}$ by
$t_f^{(k)}$ in the Schr\"odinger equation governing the dynamics of
$|\tilde{\psi}_{k+1}\rangle$. The new time $t_f^{(k+1)}$ can then be
computed from Eq.~(\ref{eqtime}) with $|\tilde{\psi}_k'\rangle$
instead of $|\tilde{\psi}_{k+1}\rangle$.

The complete iterative algorithm is described as follows:
\begin{enumerate}
\item Guess an initial control field for $k=0$ and an initial time $t_f^{(0)}$.
\item Fixing the time parameter, apply one iteration of a standard
  iterative algorithm as the one described in Sec. \ref{subsec:stand}
  to obtain $\tilde{E}_{k+1}$.
\item Fixing the control field $\tilde{E}_{k+1}$, propagate
  $|\tilde{\chi'}_k (t)\rangle$   backward
  in time  with the 'initial' condition,
  $|\tilde{\chi}(1)\rangle =|\phi_f\rangle\langle\phi_f
  |\tilde{\psi}(1)\rangle$.
  The control time is here $t_f^{(k)}$.
\item Compute the new time parameter $t_f^{(k+1)}$ from
  Eq.~(\ref{eqtime}), approximating $|\tilde{\psi}_{k+1}\rangle$ by
  $|\tilde{\psi}_k'\rangle$.
\item With the new control parameters, $\tilde{E}_{k+1}$ and
  $t_f^{(k+1)}$, go to step 2 by incrementing
the index $k$ by 1.
\end{enumerate}

%%%%%%%%%%%%%%%%%%%%%%%%%%%%%%%%%%%%%%%%%%%%%%%%%%%%%%%%%%%%%%%%%%%%%

%%%%%%%%%%%%%%%%%%%%%%%%%%%   SECTION 3   %%%%%%%%%%%%%%%%%%%%%%%%%%%

\section{The model system}\label{sec3}

We consider the control of a linear molecule by a THz laser field, linearly
polarized along the $z$-axis of the laboratory frame.
It is by now well-established that THz pulses, which interact
resonantly with molecular rotation,
induce field-free orientation and
alignment~\cite{fleischer,cong,lapert,henriksen}.

For zero rotational temperature, the dynamics of the system is
governed by the time-dependent Schr\"odinger equation,
\begin{equation}
i\frac{\partial}{\partial t}|\psi (t)\rangle =H(t) |\psi(t)\rangle ,
 \label{eq:dyn}
\end{equation}
where $H(t)$ is the Hamiltonian of the system. We use atomic units
throughout this paper unless specified otherwise.
Within the rigid rotor approximation, the Hamiltonian is given by
\begin{equation}
  H(t)=BJ^2-E(t)\mu_0\cos\theta ,
  \label{eq:ham_rot}
\end{equation}
where $B$ is the rotational constant and $\mu_0$ the permanent dipole
moment. $E(t)$ denotes the component of the electric field along the
$z$- axis. The polar angle $\theta$
is the angle between the internuclear axis and the direction of field polarization. In our
numerical examples, we consider the parameters of the CO molecule,
i.e., $B=1.9312\,$cm$^{-1}$ and $\mu_0=0.044\,$a.u. The initial state
is the ground state, denoted by $|0,0\rangle$
in the spherical harmonics basis set $\{|j,m\rangle\}$. Due to the symmetry
of the Hamiltonian with respect to the $z$- axis,
the projection $m$ of the angular momentum is a good quantum number.
This property implies that only states $|j,0\rangle$,
$j\geq 0$, will be populated during the dynamics.
In our calculations, we consider a finite Hilbert space of
size $j_{max}=15$. This size is sufficient for the intensity
of the laser field used here. For sake of simplicity, all the
numerical computations are carried out for
zero rotational temperature, but could straightforwardly be extended
to finite temperature.
%%%%%%%%%%%%%%%%%%%%%%%%%%%%%%%%%%%%%%%%%%%%%%%%%%%%%%%%%%%%%%%%%%%%%

%%%%%%%%%%%%%%%%%%%%%%%%%%%%%%%%%   SECTION 4   %%%%%%%%%%%%%%%%%%%%%%%%%%

\section{Numerical results}\label{sec4}

\subsection{State-dependent constraints in the control of molecular
  rotation}\label{sec4a}
In this section, we apply the state-dependent constraints algorithm
to maximize the molecular orientation
of the CO molecule. The terminal cost is defined as the
expectation value of the operator $\cos\theta$, $J_{t_f} =
\langle\psi(t_f)|\cos\theta|\psi(t_f)\rangle$.
This value is taken as a quantitative measure of the orientation
\cite{revuerotation2,revuerotation1}. Here,  in order to maximize both
the orientation and its duration in field-free conditions,
we add a state-dependent constraint to the optimization problem by
restricting the dynamics to the first five rotational
levels~\cite{sugnyor1,sugnyor2}, defining this to be the allowed
subspace. This choice is motivated by the fact that the duration of
field-free orientation is related, at least approximately, to the
number of states, $(j_{opt}+1)$, that make up the rotational
% on utilise j_max deja pour la taille de l'espace total
wavepacket (see Refs.~\cite{sugny,sugnyor1,sugnyor2} for mathematical
details on this relation). This time decreases when $j_{opt}$
increases. Note that a very small subspace with $j_{opt}=4,5$ turns
out to be a good compromise for efficient and long-lived orientation
\cite{sugny}.  We measure the population in the
allowed subspace by the temporal average
\begin{equation}
  I_{p} = \frac{1}{t_f}\int_0^{t_f} \langle \psi(t)| \hat{P}
  |\psi(t)\rangle dt\,,
  \label{eq:Ip}
\end{equation}
where $\hat{P}$ is the projector onto the space $\{|j,0\rangle\}$,
$j=0,\cdots, 4$.
The guess field is taken to be a Gaussian pulse of
144$\,$fs full width at half maximum (FWHM), centered at $t_0 =
T_{\textrm{per}}/5$, $T_{\textrm{per}}$ being the rotational period of the molecule.
The control time $t_f$ is chosen to be equal to $T_{\textrm{per}}$.

\begin{figure}[tb]
  \centering
  \includegraphics[width=0.9\linewidth]{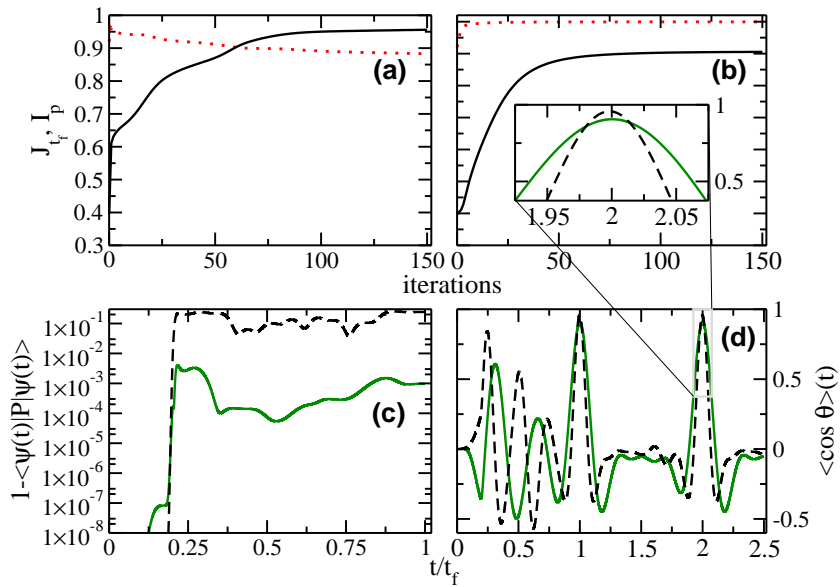}
  \caption{(Color online)
    Terminal cost $J_{t_f}$ (solid black line) and average
    population of the allowed subspace $I_p$ (dashed red
    line)  as a function of the number of iterations for
    standard optimization (a) and including the state-dependent
    constraint (b). The corresponding time evolution of the population in the
    forbidden subspace, $1-\langle \psi(t)| \hat{P} |\psi(t)\rangle$,
    is shown in panel (c), with (solid green line) and without (dashed
    black line) state-dependent
    constraint. The time evolution of the figure of merit for orientation,
    the  expectation value of $\cos\theta$, is displayed in
    panel (d) for fields optimized with (solid green line) and
    without (dashed black line) the state-dependent constraint. The
    parameters $\lambda$ and $\mu$ used in the simulations are fixed
    to 20 and 50/$t_f$, respectively. The small insert in panel (b) represents a zoom of the
    dynamical evolution near a maximum of $\langle\cos\theta\rangle$.}
  \label{fig:JT_sdc}
\end{figure}
Figure~\ref{fig:JT_sdc} compares the results obtained by a standard
optimization procedure and by one employing the state-dependent
constraint. The fidelities achieved by the optimized fields
are about 95$\%$ and 91$\%$ for the standard and state-dependent
constraint optimization algorithms, respectively.
The optimized field without the constraint leads to
a larger orientation due to population of states $j>j_{opt}$. This is
evident from inspection of  the dashed black curve in
Fig. \ref{fig:JT_sdc}(c) and corresponds to exploring
solutions in the forbidden subspace.
At the end of the optimization process, more than 25$\%$ of the population
remains in the forbidden subspace. The solid green curve of
Fig.~\ref{fig:JT_sdc}(c) demonstrates that population stays
almost completely within the allowed subspace if the state constraint
is taken into account. The population transfer towards the forbidden
subspace is reduced by two orders of magnitude.
Figure~\ref{fig:JT_sdc}(d) shows the time evolution of
$\langle\cos\theta\rangle$ during two and a half rotational
periods. During the first rotational period the field is on,
afterwards the evolution is field-free.
The optimized field without the state-dependent constraint achieves
a higher molecular orientation than the one obtained with the
constraint also in the field-free case, due to the rotational
wavepacket being made up of states with high $j$.
However, as could be expected, the constraint formulation
leads to a longer duration of the
orientation in field-free conditions. More precisely, we get a FWHM of
0.086 and 0.136 without and with constraint. The duration of the
field-free orientation amounts to what can be expected for a
rotational wavepacket being comprised of $j\le\j_{opt}$~\cite{sugny}.

\subsection{Time-optimization of molecular rotation}
\label{sec4b}
This section focuses on the production of molecular orientation by
a joint optimization of the control time and of the
laser fluence. Instead of optimizing the expectation
value $|\langle \cos\theta\rangle|$ itself, we choose here a target
state $|\phi_f\rangle$ which maximizes
$|\langle \cos\theta\rangle|$ in a finite-dimensional Hilbert space,
$\mathcal{H}_{j_{opt}}$, spanned by the states $\{|j,0\rangle\}$ with
$0\leq j\leq j_{opt}$. The details of the construction of
$|\phi_f\rangle$ are found in Refs.~\cite{sugnyor1,sugnyor2}. In the numerical computations,
the parameter $j_{opt}$ is fixed to the value 4. The control scheme is
the one described in Sec.~\ref{subsec:to} with
the guess field of Sec. \ref{sec4a}.
%The numerical results are displayed in Fig.~\ref{fig:JT_to}.
\begin{figure}[tb]
\centering
\includegraphics[width=0.9\linewidth]{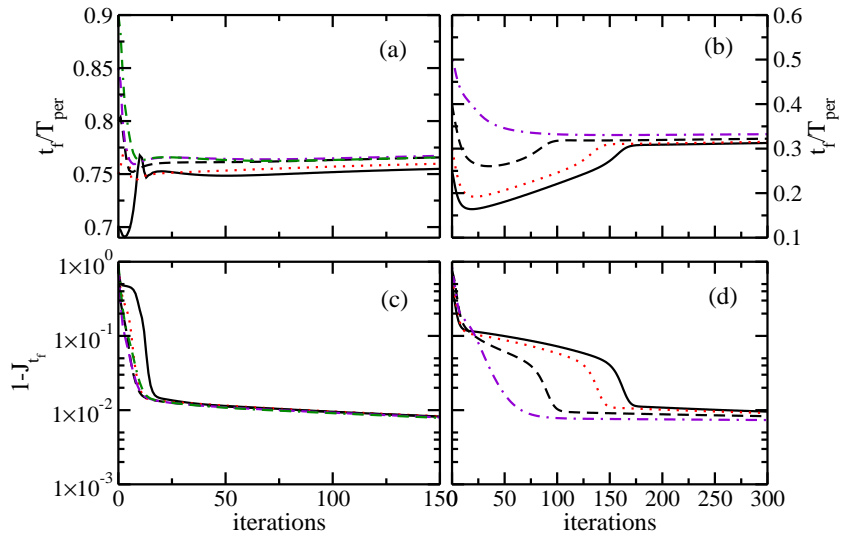}
\caption{(Color online)
  Normalized control time $t_f/T_{\mathrm{per}}$ as a function of the
  number of iterations for different values of $t_f^{(0)}$ (panels (a)
  and (b)). Deviation from unity of the corresponding terminal cost
  (panels (c) and (d)).
  The parameters $\lambda$ and $\epsilon$ are fixed to 5 and 1000 respectively.
}
\label{fig:JT_to}
\end{figure}
Figure~\ref{fig:JT_to}(a) and (b) illustrates that $t_f$ is indeed
independent of the initial time $t_f^{(0)}$, a crucial property of the
algorithm. Two limit points of the sequence $(t_f^{(k)})$ are found for
$t_f^0/T_\mathrm{per}\in [0.25,0.9]$.
The two attraction points are  $t_f/T_\mathrm{per}
\approx 0.31$ for $t_f^{(0)}/T_\mathrm{per} \in [0.25,0.6]$
and $t_f/T_\mathrm{per} \approx 0.77$ for $t_f^{(0)}/T_\mathrm{per} \in [0.75,0.9]$.
Note that other limit points can be found if
$t_f^{(0)}>T_\mathrm{per}$. Extensive numerical tests show
that the two attraction points are not changed by small modifications
of $\lambda$ and the maximum amplitude of the guess field. The
optimal solutions corresponding to the two limit points, shown in
Fig.~\ref{fig:JT_to}(c) and (d), have a very similar efficiency.

The numerical algorithm has revealed that two times,
clearly shorter than one rotational period, are well-suited to
maximize molecular orientation.
This result is interesting in view of practical applications since, in
standard experimental conditions~\cite{seideman,vieillard},
collisions play a significant role only for
times larger than 3 or 4 rotational periods.

%%%%%%%%%%%%%%%%%%%%%%%%%%%%%%%%%%%%%%%%%%%%%%%%%%%%%%%%%%%%%%%%%%%%%%%%%%%
\section{Conclusion}\label{sec5}
We have applied two recent formulations of optimal control
algorithms to the manipulation of molecular orientation by THz laser
fields. Such algorithms have the advantage of simplicity and general
applicability for any quantum dynamics. The molecular rotation studied
here serves as an illustrative example to demonstrate the efficiency
of the methods.

Moreover, our work provides insight into the different
ways to produce molecular orientation. For this complicated control
problem, there exists no unique optimal solution. A particular pathway
can be selected by adding constraints, either on the control field (here
the optimization of the control time) or on the state space (here only
a subspace of the total Hilbert space is allowed to be populated by
the control field). Generally, constraints could also be designed to
account for experimental imperfections or requirements
related to a specific material or device.
The possibility of including such constraints
renders the optimal control theory more useful in view of
experimental applications and helps bridge the gap between control
theory and control experiments.

%%%%%%%%%%%%%%%%%%%%%%%%%%%%%%%%%%%%%%%%%%%%%%%%%%%%%%%%%%%%%%%%%%%%%%%%%%%%%

%%%%%%%%%%%%%%%%%%%%%%%%%%%   CONCLUSION   %%%%%%%%%%%%%%%%%%%%%%%%%%%%%%%%%%

%\section{Conclusions}\label{sec5}

%%% No conclusions?

%%%%%%%%%%%%%%%%%%%%%%%%%%%%%%%%%%%%%%%%%%%%%%%%%%%%%%%%%%%%%%%%%%%%%%%%%%%%%

\section{Acknowledgments}

We gratefully acknowledge financial support from the Conseil R\'egional de Bourgogne and the QUAINT coordination action (EC FET-Open).

\end{document}